\documentclass[aps,pra,preprint]{revtex4}
\usepackage{graphicx}
\usepackage{dcolumn}
\usepackage{amsmath}
\usepackage{amssymb}
\begin{document}
\title{On the ``Causality Paradox" of Time-Dependent Density Functional Theory} 
\author{Giovanni Vignale}
\affiliation{Department of Physics, University of Missouri-Columbia,
Columbia, Missouri 65211}
\date{\today}
\begin{abstract}
I show that the so-called causality paradox of time-dependent density functional theory arises from an incorrect formulation of the variational principle for the time evolution of the density.  The correct formulation not only resolves the paradox in real time, but also leads to a new expression for the causal exchange-correlation kernel in terms of Berry curvature.  Furthermore, I show that all the results that were previously derived  from symmetries of the action functional remain valid in the present formulation.  Finally, I develop a model functional theory which explicitly demonstrates the workings of the new formulation.
\par
\end{abstract}
\maketitle
\section{Introduction} 
Time-dependent density functional theory (TDDFT)~\cite{TDFTBook,Runge84,Gross90}  is becoming a standard tool for the computation of time-dependent phenomena in condensed matter physics and quantum chemistry.  Naturally the growing number of application has generated a new interest in the foundations of the theory (see, for example the recent critique by Schirmer and Drew~\cite{Schirmer07}, and the rebuttal by Maitra, Burke, and van Leeuwen~\cite{Maitra07}).  In this paper I address the so-called ``causality paradox", a problem that has troubled TDDFT for many years~\cite{Gross96}, and has been the object of many discussions and technically sophisticated resolutions~\cite{Rajagopal96,vanLeeuwen98,vanLeeuwen01,Mukamel05,TDFTBook}.  I do not disagree with those resolutions, but I wish to propose a new one, which I find technically simpler, more direct, and closer to the spirit of the original formulation of TDDFT.  

As an introduction to the problem, let us recall that the formal basis of TDDFT is the Runge-Gross (RG) theorem\cite{Runge84}, which establishes a biunivocal correspondence between the time dependent particle density  $n({\bf r},t)$ of a many-body system and the  potential $v({\bf r}, t)$ that gives rise to that density starting from an assigned quantum state $|\psi_0\rangle$ at the initial time $t=0$.    According to the RG theorem the potential that gives rise to $n({\bf r},t)$ starting from $|\psi_0\rangle$ is determined by $n({\bf r},t)$ and $|\psi_0\rangle$ up to a time-dependent constant. Similarly, the time-dependent quantum state $|\psi(t)\rangle$ is determined by $n({\bf r},t)$ and $|\psi_0\rangle$  up to a time-dependent phase factor.  In this sense, both $v({\bf r},t)$ and $|\psi(t) \rangle$ are {\it functionals} of $n({\bf r},t)$ and $|\psi_0\rangle$ over a time interval $0 \leq t \leq T$.  They should be denoted by  $v[n,|\psi_0\rangle;{\bf r},t]$ and $|\psi[n,|\psi_0\rangle]\rangle$ respectively.  From now on, however, the dependence on the initial state will not be explicitly noted and we will write simply $v[n;{\bf r},t]$ and $|\psi[n]\rangle$.    Admittedly, there is no proof that every reasonable density can be produced by some local potential: but it is generally assumed that the densities for which the theorem holds are dense enough in the space of densities to provide an arbitrarily good approximation to the physical densities one might encounter in real life.  In this paper I will assume {\it tout court} that all the time-dependent densities can be produced by some local potential: i.e., all time-dependent densities are {\it v-representable}.

Another theorem, proved by van Leeuwen in 1999,\cite{vanLeeuwen99} extends and strengthens the RG theorem.  According to van Leeuwen's theorem, the density  $n({\bf r},t)$ which evolves in an interacting system  under the action of an external potential $v({\bf r},t)$ starting from an initial state $|\psi_0\rangle$, can be reproduced in a noninteracting system evolving under the action of an appropriate and uniquely determined potential $v_s({\bf r},t)$, starting from any initial state  $|\psi_{s0}\rangle$ that has the same density and divergence of the current density as $|\psi_0\rangle$.  This theorem provides the basis for the extremely useful Kohn-Sham method of calculating the density.    The effective potential $v_s({\bf r},t)$ -- a functional of $n({\bf r},t)$, $|\psi_0\rangle$, and $|\psi_{s0}\rangle$ -- is known as the Kohn-Sham potential.  The difference $v({\bf r},t) - v_s({\bf r},t)-v_H({\bf r},t)$, where $v_H({\bf r},t)$ is the Hartree potential, is known as the exchange-correlation (xc) potential, denoted by $v_{xc}[n;{\bf r},t]$ -- also a functional of $n({\bf r},t)$, $|\psi_0\rangle$, and $|\psi_{s0}\rangle$.  

In general the potentials $v({\bf r},t)$, $v_s({\bf r},t)$ and $v_{xc}({\bf r},t)$ depend on the density $n({\bf r}',t')$ at different positions and earlier times $t'<t$, but cannot be affected by changes in the density at later times $t'>t$.  This obvious  causality requirement implies that  the functional derivatives of  these potentials with respect to $n({\bf r},t')$ and, in particular, the {\it exchange-correlation kernel} $f_{xc}({\bf r},t;{\bf r}',t') \equiv \delta v_{xc}[n;{\bf r},t]/\delta n({\bf r}',t')$ vanish for $t<t'$.  

An interesting question is whether the potentials  $v({\bf r},t)$, $v_s({\bf r},t)$ and $v_{xc}({\bf r},t)$ can be generated from functional derivatives of an action functional $A[n,|\psi_0\rangle]$ (denoted from now on simply as $A[n]$)  with respect to the density, in close analogy with  static DFT, where the potentials are functional derivatives of energy functionals with respect to the density.  The existence of such a representation was suggested by RG in their original paper~\cite{Runge84}, and was subsequently used by this author~\cite{Vignale95,Vignale96} to derive several theorems in TDFT.  In the mid-nineties, however, it became clear that the representation was problematic to say the least~\cite{Gross96}.  If the potential could be written as a functional derivative of an action functional, 
\begin{equation}\label{RGV}
v[n;{\bf r},t]\equiv\frac{\delta A[n]}{\delta n({\bf r},t)}~,
\end{equation}
then we should also have 
\begin{equation}
\frac{\delta v[n;{\bf r},t]}{\delta n({\bf r}',t')} =  \frac{\delta^2 A[n]}{\delta n({\bf r},t)\delta n({\bf r}',t')}~.
\end{equation}
But this equation is patently false, because the left hand side is different from zero only for $t>t'$ (by the causality requirement), while the right hand side is symmetric under interchange of $t$ and $t'$.

This startling observation became quickly known as {\it the causality paradox} and prompted several sophisticated resolutions~\cite{Rajagopal96,vanLeeuwen98,vanLeeuwen01,Mukamel05}. The best known is the Van Leeuwen's construction of a ``Keldysh action" in pseudotime~\cite{vanLeeuwen98,vanLeeuwen01}.   More recently Mukamel~\cite{Mukamel05} has shown how to construct causal response functions  from symmetrical functional derivatives corresponding to ``Liouville space pathways".   The gist of these resolutions is that causality is not violated, but one must use a more abstract mathematical apparatus (Keldysh formalism, or the Liouville superoperator method) in order to connect  functional derivatives of the action to causal response functions.

In this paper I re-examine the ``paradox" from a more elementary point of view.  I show that the variational principle for the time-evolution of the wave function, when properly implemented as a variational principle for the density,  yields an expression for the potential as the sum of two terms: (1) the functional derivative of the RG action and (2) a correction term, which cannot be expressed as a functional derivative, but is still simple enough to be included in all the formal proofs.   So the gist of the present resolution is that we learn to write the potential as functional derivative of an action {\it plus} a boundary term.  Among other benefits, this approach explains why theorems that were originally proved under the incorrect assumption~(\ref{RGV})\cite{Vignale95,Vignale96}, turned out to be true after all.  Furthermore, it leads to interesting expressions for the inverse of the density-density response function and the xc kernel in terms of ``Berry curvature".

This paper is organized as follows.  In the next section I discuss in detail the failure of the stationary action principle for the density, and show how the correct causal expressions for the xc potential and the xc kernel are derived from a modified variational principle.  In section~\ref{Theorems} I explain why in many cases one can still pretend that the xc potential is the functional derivative of the xc action and get correct results.  Appendix~\ref{ETS} clears up a technical point about the equal-time singularities of causal response functions.  Finally, Appendix~\ref{Model} presents a pedagogical ``time-dependent position density functional theory", which is conceptually equivalent to the full-fledged TDDFT but can be solved exactly, illustrating the workings of the new formulation.  
%

\section{Variational principle for the density} \label{section2}
The starting point is the time-dependent quantum variational principle,\cite{Saraceno81} according to which the time-dependent Schr\"odinger equation is equivalent to the requirement that the action
\begin{equation}\label{Frenkel}
A_V[|\psi\rangle] = \int_0^T \langle \psi(t)|i \partial_t - \hat H_V|\psi(t)\rangle  dt 
\end{equation}
($\hbar =1$) be stationary ($\delta A_V=0$) with respect to a  arbitrary variations of the wave function which vanish at the ends of the time interval $0\leq t\leq T$, i.e. $|\delta\psi(0)\rangle=|\delta \psi(T)\rangle=0$.  Here 
\begin{equation}\label{def-HV}
\hat H_V=\hat H_0+\int V({\bf r},t) \hat n({\bf r}) d{\bf r}
\end{equation}
is the sum of the internal hamiltonian $\hat H_0$ (kinetic + potential) and the interaction with an external time-dependent potential field $V({\bf r},t)$.   

The proof is straightforward.  The variation of $A_V$ induced by a variation $\delta|\psi\rangle \equiv |\delta\psi \rangle$ is
\begin{eqnarray}\label{variation1}
\delta A_V[|\psi\rangle] &=&  \int_0^T \langle \delta \psi(t)|i \partial_t - \hat H_V|\psi(t)\rangle  dt  \nonumber\\ &+&  \int_0^T \langle \psi(t)|i \partial_t - \hat H_V|\delta\psi(t)\rangle  dt~,
\end{eqnarray}
and the second term on the right hand side can be integrated by parts to yield
\begin{eqnarray}\label{variation2}
\delta A_V[|\psi\rangle] &=&  \int_0^T \langle \delta \psi(t)|i \partial_t - \hat H_V|\psi(t)\rangle  dt  \nonumber\\ &+&  \int_0^T \langle (i \partial_t - \hat H_V)\psi(t)|\delta\psi(t)\rangle  dt \nonumber\\
&+&i\left.\langle \psi(t)|\delta\psi(t)\rangle\right\vert_{t=0}^{t=T}~.
\end{eqnarray}
The last term on the right hand side vanishes by virtue of the boundary conditions on $|\delta \psi(t)\rangle$, and the vanishing of the first two terms is equivalent to the time-dependent Schr\"odinger equation  $(i\partial_t - \hat H_V)|\psi(t)\rangle=0$.

Since $|\psi(t)\rangle$ is, by virtue of the RG theorem, a functional of $n$ and $|\psi_0\rangle$,  Runge and Gross suggested that a stationary action principle for the density could be formulated in terms of the functional
\begin{eqnarray}\label{RGVAction}
A_V[n] &=& \int_0^T \langle \psi[n]|i \partial_t - \hat H|\psi[n]\rangle dt \nonumber\\
&=&A_0 [n] - \int_0^T V({\bf r},t)n({\bf r},t) d{\bf r} dt~,
\nonumber\\
\end{eqnarray}
where the ``internal action"  
\begin{equation}\label{RGAction}
A_0 [n]\equiv \int_0^T \langle \psi[n]|i \partial_t - \hat H_0|\psi[n]\rangle dt 
\end{equation}
 is a universal functional of the density and the initial state.  
 Then, setting $\delta A_V = 0$ for arbitrary variations of the density we easily find 
\begin{equation}
V({\bf r},t)=\frac{\delta A_0[n]}{\delta n({\bf r},t)},
\end{equation}
when $n({\bf r},t)$ is the density corresponding to $V({\bf r},t)$.  This implies that the external potential, {\it viewed as a functional of the density},  is the functional derivative of the internal action with respect to the density:
\begin{equation}\label{v-functional}
v[n;{\bf r},t]\equiv\frac{\delta A_0[n]}{\delta n({\bf r},t)},
\end{equation}
and the time evolution of the density is determined by requiring $v[n;{\bf r},t]=V({\bf r},t)$, where $V({\bf r},t)$ is the actual external potential.
The only problem with Eq.~(\ref{v-functional}), which would otherwise be very useful, is that it  plainly contradicts causality, as discussed in the introduction.  What went wrong?

The problem arises from the fact that the Frenkel variational principle $\delta A_V=0$ is valid only for variations of $|\psi\rangle$ that vanish at the endpoints of the time interval under consideration, i.e. at $t=0$ and $t=T$.  But a variation of the density at any time $t<T$  inevitably causes a change in the quantum state at time $T$.  Therefore we can only set $|\delta\psi(0)\rangle =0$, but have no right to set  $|\delta \psi(T)\rangle =0$.   Taking this into account, and going back to Eq.~(\ref{variation2}) we see that the correct formulation of the variational principle for the density is not $\delta A_V=0$ but
\begin{equation}\label{pre-main}
\delta A_V[n] =  i\langle \psi_T[n]|\delta \psi_T[n]\rangle~. 
\end{equation}
Here $|\psi_T[n]\rangle \equiv|\psi[n;T]\rangle$ is the quantum state at time $T$ regarded as a functional of the density (and of course of the initial state).    So we see that the action functional is not stationary, but its variation must be equal to another functional of the density, which is given on the right hand side of Eq.~(\ref{pre-main}).

Taking the functional derivative of Eq.~(\ref{pre-main}) with respect to $n({\bf r},t)$ and making use of Eqs.~(\ref{RGVAction}) and~(\ref{v-functional})  we get
\begin{equation}\label{mainresult}
v[n;{\bf r},t] = \frac{\delta A_0[n]}{\delta n({\bf r},t)}-i\left \langle \psi_T[n] \left\vert \frac{\delta \psi_T[n]}{\delta n({\bf r},t)}\right\rangle \right.~, 
\end{equation}
where $\left.\left\vert \frac{\delta \psi_T[n]}{\delta n({\bf r},t)}\right\rangle \right.$ is a compact representation for the functional derivative of $|\psi_T[n]\rangle$ with respect to density.

This is the main result of this paper, since it shows that the external potential (and hence also the Kohn-Sham potential and the xc potential) is not merely a functional derivative of the Runge-Gross action $A_0[n]$.  Notice that the additional ``boundary term" is real, in spite of the $i$,  because the quantum state $|\psi_T[n]\rangle$ is normalized to $1$ independent of density, implying that
\begin{equation}
\left \langle \psi_T[n] \left\vert \frac{\delta \psi_T[n]}{\delta n({\bf r},t)}\right\rangle \right. = 
-\left.\left \langle \frac{\delta \psi_T[n]}{\delta n({\bf r},t)}\right\vert  \psi_T[n]\right\rangle
\end{equation}
is a purely imaginary quantity.

 At first sight, however, Eq.~(\ref{mainresult}) is still problematic because it appears to depend on the arbitrary upper limit of the time interval ($T$) and therefore also on the density at times $t'>t$.    However, this is only appearance.  The point is that both the functional derivative of $A_0[n]$ and the boundary term, considered separately,  have a non-causal dependence on the density, but the dependence on $n(t')$ with $t'>t$ cancels out exactly when the two terms are combined!   
 
 Let us show this in detail.   Consider, for example, an increment of the upper limit of the time interval from $T$ to $T+\Delta T$.  The density $n({\bf r},t)$ must be smoothly continued to the larger time interval $[0,T+\Delta T]$, and the quantum state $|\psi[n]\rangle$ satisfies in this time interval the Schr\"odinger equation 
\begin{equation} \label{SE1}
 (i\partial _t  - \hat H_0 - \hat v[n;t])|\psi[n;t]\rangle=0~,
 \end{equation}
where   
\begin{equation}
\hat v[n;t] \equiv \int v[n;{\bf r}',t]\hat n({\bf r}') d{\bf r}'~, 
\end{equation}
$\hat n({\bf r})$ is the density {\it operator}, and $v[n;{\bf r},t]$ is the potential that yields $n({\bf r},t)$.  
Now  in view of Eq.~(\ref{SE1}) the internal action over the time-domain $[0,T]$ can be written as 
\begin{equation}
A_0[n]=\int_0^{T} \langle \psi[n;t']|\hat v[n;t'])|\psi[n;t']\rangle dt'~,
\end{equation}
and its variation, due to the extension of the upper limit from $T$ to $T+\Delta T$ is, to first order in $\Delta T$ given by
\begin{equation}
\Delta A_0[n]=\Delta T \langle \psi_T[n]|\hat v[n;T]|\psi_T[n]\rangle~.
\end{equation}
Taking the functional derivative with respect to $n({\bf r},t)$, with $t$ within the interval $[0,T]$,
we see that
\begin{equation}\label{Proof-step1}
\Delta \frac{\delta A_0[n]}{\delta n({\bf r},t)}=\Delta T \langle \psi_T[n]\left \vert \frac{\delta \hat v[n;T]}{\delta n({\bf r},t)}\right\vert\psi_T[n]\rangle~.
\end{equation}
The reason why we could take the functional derivative inside the expectation value on the right hand side of this equation is that $\langle \psi_T[n]|\hat v[n;T]|\psi_T[n]\rangle = \int v[n;{\bf r}',T] n({\bf r}',T) d{\bf r}'$ depends on $n({\bf r},t)$ only through the potential functional $v[n;{\bf r}',T]$: the density $n({\bf r}',T)$ is, by definition, unaffected by a variation of the density at the earlier time $t$.\footnote{More formally, $\frac{\delta n({\bf r},T)}{\delta n({\bf r}',t)} = \delta ({\bf r} - {\bf r}') \delta(T-t)$, which vanishes if $T>t$.  It should be noted that the nature of our variational principle for the density is such that we are allowed to impose the condition $\delta n({\bf r},t) =0$ at $t=0$ and $t=T$, just as in the formulation of the variational principle for the wave function we can assume that the variations of the wave function vanish at $t=0$ and $t=T$. This makes the derivation of Eq.~(\ref{Proof-step1}) even simpler.}  

Consider now the variation of the boundary term of Eq.~(\ref{mainresult}) again due to the change of the upper limit of the time interval from $T$ to $T+\Delta T$.  To first order in $\Delta T$ we have
\begin{equation}
|\psi_{T+\Delta T}[n]\rangle -|\psi_{T}[n]\rangle  = - i \Delta T \hat H[n;t]|\psi_{T}[n]\rangle~,
\end{equation}
where $\hat H[n,t]= \hat H_0 +\hat v[n;T]$ is the full time-dependent hamiltonian regarded as a functional of the density.
Substituting this in the variation of the boundary term we get
\begin{eqnarray}
&&-i \Delta \left \langle \psi_T[n] \left\vert \frac{\delta \psi_T[n]}{\delta n({\bf r},t)}\right\rangle \right. =\nonumber\\ && \Delta T
\left\langle \psi_T[n] \left \vert \hat H[n;T]\frac{\delta}{\delta n({\bf r},t)} - \frac{\delta}{\delta n({\bf r},t)}\hat H[n;T] \right\vert \psi_T[n]\right \rangle.\nonumber\\
\end{eqnarray}

This seemingly complicated expression contains the commutator between the Hamiltonian  $\hat H[n;T]$ and the functional derivative with respect to $n({\bf r},t)$.  This commutator is simply $- \frac{\delta \hat H[n;T]}{\delta n({\bf r},t)}$, which is evidently equal to $ - \frac{\delta \hat v[n;T]}{\delta n({\bf r},t)}$.  Thus the variation of the boundary term is given by
\begin{equation}\label{Proof-step2}
-i \Delta \left \langle \psi_T[n] \left\vert \frac{\delta \psi_T[n]}{\delta n({\bf r},t)}\right\rangle \right. = -\Delta T
\left\langle \psi_T[n] \left\vert \frac{\delta \hat v[n;T]}{\delta n({\bf r},t)} \right\vert \psi_T[n]\right \rangle.
\end{equation}
Combining the two equations~(\ref{Proof-step1}) and ~(\ref{Proof-step2}) we see that net variation of the potential $v[n;{\bf r},t]$ is exactly zero.  This means that we have the freedom to change at will the upper limit $T$ of the time interval: the value of  $v[n;{\bf r},t]$  will not change, as long as $T$ remains larger than or equal to $t$.  But this means that we can always choose $T=t$, and this proves that the potential at time $t$ does not depend on what happens to the density at times later than $t$.  QED. 


Now that we have the correct expression for the potential as a functional of the density it is easy to construct the other two potentials of interest, namely the Kohn-Sham potential and the xc potential.
For the Kohn-Sham potential we simply have
\begin{equation}\label{mainresultKS}
v_s[n;{\bf r},t] = \frac{\delta A_{0s}[n]}{\delta n({\bf r},t)}-i\left \langle \psi_{sT}[n] \left\vert \frac{\delta \psi_{sT}[n]}{\delta n({\bf r},t)}\right\rangle \right.~, 
\end{equation}
where $A_{0s}$ is the internal action for a noninteracting system and $|\psi_{sT}[n]\rangle$ is the non-interacting version of $|\psi_T[n]\rangle$ (starting from an initial state $|\psi_{s0}\rangle$).  For the exchange-correlation potential we get
\begin{eqnarray}\label{mainresultxc}
&&v_{xc}[n;{\bf r},t] = \frac{\delta A_{xc}[n]}{\delta n({\bf r},t)}\nonumber\\&&-i\left \langle \psi_T[n] \left\vert \frac{\delta \psi_T[n]}{\delta n({\bf r},t)}\right\rangle \right.+i\left \langle \psi_{sT}[n] \left\vert \frac{\delta \psi_{sT}[n]}{\delta n({\bf r},t)}\right\rangle \right., \nonumber\\
\end{eqnarray}
where $A_{xc}[n]$ is the usual xc action defined in the Runge-Gross paper as the difference $A_0[n]-A_{0s}[n]-A_H[n]$, where the last term is the Hartree action.  Notice that $v_{xc}$ depends not only on the density, but also on the two initial states $|\psi_0\rangle$ and $|\psi_{s0}\rangle$.

The above formulas allow us to obtain an elegant expression for the functional derivative of the potential with respect to the density.  Consider, for example, the functional derivative of $v(n;{\bf r},t)$.  From Eq.~(\ref{mainresult}) we see that this is given by

\begin{eqnarray}\label{result1}
\frac{\delta v[n;{\bf r},t]}{\delta n({\bf r}',t')}= \frac{\delta^2 A_0[n]}{\delta n({\bf r},t)\delta n({\bf r}',t')}-i\left \langle \psi_T[n]\left\vert\frac{\delta^2 \psi_T[n]}{\delta n({\bf r},t)\delta n({\bf r}',t')}\right\rangle \right.
 -i\left \langle \frac{\delta\psi_T[n]}{\delta n ({\bf r}',t')}\left\vert \frac{\delta \psi_T[n]}{\delta n({\bf r},t)}\right\rangle\right.~.
\end{eqnarray}

The first two terms on the right hand side are symmetric under interchange of ${\bf r},t$ and ${\bf r}',t'$, while the last term, which has the structure of a {\it Berry curvature}, is antisymmetric under the same interchange.  Let us then subtract from Eq.~(\ref{result1}) the same equation with ${\bf r},t$ and ${\bf r}',t'$ interchanged.  When $t>t'$ then $\frac{\delta v[n;{\bf r}',t']}{\delta n({\bf r},t)}$ vanishes because of causality, so on the left hand side only $\frac{\delta v[n;{\bf r},t]}{\delta n({\bf r}',t')}$ is left.   And on the right hand side the symmetric terms cancel out, leaving only the Berry curvature term.  The result is
\begin{eqnarray}\label{result2}
\frac{\delta v[n;{\bf r},t]}{\delta n({\bf r}',t')} &=& -i\left[\left \langle \frac{\delta\psi_T[n]}{\delta n ({\bf r}',t')}\left\vert \frac{\delta \psi_T[n]}{\delta n({\bf r},t)}\right\rangle \right.
-\left \langle \frac{\delta\psi_T[n]}{\delta n ({\bf r},t)}\left\vert \frac{\delta \psi_T[n]}{\delta n({\bf r}',t')}\right\rangle \right. \right] \nonumber\\ 
&=&2  \Im m \left \langle \frac{\delta\psi_T[n]}{\delta n ({\bf r}',t')}\left\vert \frac{\delta \psi_T[n]}{\delta n({\bf r},t)}\right\rangle \right.~~~~~~(t>t').
\end{eqnarray}
Notice that the right hand side of this formula is independent of $T$ (as long as $T$ is larger that $t$ and $t'$) since, as we have shown, $v[n;{\bf r},t]$ satisfies the causality requirements.  

The above argument determines $\frac{\delta v[n;{\bf r},t]}{\delta n({\bf r}',t')}$ for $t>t'$, but leaves open the possibility of singular contributions at $t=t'$.   
In Appendix~\ref{ETS} I show that indeed the functional derivative $\delta v[n;{\bf r},t]/\delta n({\bf r}',t')$ contains an equal-time singularity of the form
\begin{equation}
C_0 \delta(t-t')+C_1\dot\delta(t-t')+C_2 \ddot\delta(t-t'),
\end{equation}
where $C_0$, $C_1$, and $C_2$ are functionals of the density at time $t$ ($\equiv n_t$) and functions of ${\bf r}$ and ${\bf r}'$.  $\dot \delta$ and $\ddot \delta$ denote, respectively, the first and the second derivative of the $\delta$-function with respect to its own argument.  Thus the complete expression for $\delta v/\delta n$ has the following form:
\begin{eqnarray}\label{result3}
\frac{\delta v[n;{\bf r},t]}{\delta n({\bf r}',t')} = 2 \theta(t-t')  \Im m \left \langle \frac{\delta\psi_T[n]}{\delta n ({\bf r}',t')}\left\vert \frac{\delta \psi_T[n]}{\delta n({\bf r},t)}\right\rangle \right. + \hat S_\infty[n_t;{\bf r},{\bf r}']\delta(t-t')~,
\end{eqnarray}
where $\hat S_\infty[n_t;{\bf r},{\bf r}']$ is the differential operator $C_0+C_1\frac{d}{dt}+C_2\frac{d^2}{dt^2}$.    The coefficient of the leading term, $C_2$, is independent of interactions, and the coefficient of the linear term, $C_1$, vanishes in the linear response limit.   An explicit demonstration of the equal-time singularities is provided in  Eq.~(\ref{f-derivative-singular}) of Appendix~\ref{Model}.


Equal-time singularities also enter the expression of the exchange-correlation kernel.  Taking into account the fact that $C_2$ is independent of interactions we find 
\begin{eqnarray}\label{result3-xc}
f_{xc}[n;{\bf r},t,{\bf r}',t'] &\equiv& \frac{\delta v_{xc}[n;{\bf r},t]}{\delta n({\bf r}',t')} \nonumber\\&=& 2 \theta(t-t') \left\{\Im m \left \langle \frac{\delta\psi_T[n]}{\delta n ({\bf r}',t')}\left\vert \frac{\delta \psi_T[n]}{\delta n({\bf r},t)}\right\rangle \right. - \Im m \left \langle \frac{\delta\psi_{sT}[n]}{\delta n ({\bf r}',t')}\left\vert \frac{\delta \psi_{sT}[n]}{\delta n({\bf r},t)}\right\rangle \right.\right\}\nonumber\\&+&\Delta C_1[n_t;{\bf r},{\bf r}'] \dot \delta (t-t')+f_{xc,\infty}[n_t;{\bf r},{\bf r}']\delta(t-t'),\nonumber\\
\end{eqnarray}
where $\Delta C_1$ is the difference between the coefficients $C_1$ in the interacting and non-interacting systems and  $f_{xc,\infty}$ denotes the difference between the coefficients $C_0$ in the interacting and non-interacting system.  In the linear response regime, i.e. when the time-dependent potential is a weak perturbation to the ground-state, $\Delta C_1$ vanishes and $f_{xc,\infty}$ reduces to the well-known infinite-frequency xc kernel of linear response theory.\cite{TheBook}  This behavior is demonstrated in Appendix~\ref{Model} for our exactly solved model -- see Eq.~(\ref{fxc-derivative-singular}).


Finally,  I note that the adiabatic approximation to the xc kernel is given by
\begin{equation}
f_{xc}^{ad}({\bf r},t,{\bf r}',t')  \simeq f_{xc,0}({\bf r},{\bf r}',t) \delta(t-t')~,
\end{equation}
where  
\begin{equation}
f_{xc,0}({\bf r},{\bf r}',t) =f_{xc,\infty}[n_t;{\bf r},{\bf r}']+\int_0^t  f_{xc}[n;{\bf r},t,{\bf r}',t']dt'
\end{equation} 
is the integral of the exchange-correlation kernel over all times $t'$ earlier than $t$.  The implicit assumption here is that the retardation range of the $xc$ kernel is much shorter than the time scale of variation of the density, so that the $xc$ kernel can effectively be approximated as a $\delta$-function on that time scale.  The first term on the right hand side of this expression  is the contribution of the ``true" $\delta$-function terms of Eq.~(\ref{result3-xc}).  The time integral in the second term is restricted to times strictly less than $t$.

\section{Why did ``$v_{xc}[n;{\bf r},t]=\frac{\delta A_{xc}[n]}{\delta n({\bf r},t)}$" work?}\label{Theorems}
The incorrect representation of the xc potential as a functional derivative of the xc action played a significant role in the early development of TDDFT, particularly in the proof of theorems that depend on symmetries of the action functional.  Consider, for example, the ``zero-force theorem"~\cite{Vignale95,Vignale96}, according to which the net force exerted by the xc potential on the system is zero.  This theorem was originally derived from the apparent invariance of the $xc$ action under a homogeneous time-dependent translation of the density:
\begin{equation}\label{Axc-invariance}
A_{xc}[n']=A_{xc}[n]
\end{equation}
where $n'({\bf r},t) = n({\bf r}+ {\bf x}(t),t)$, and ${\bf x}(t)$ is an arbitrary time-dependent displacement that vanishes at $t=0$.  The invariance of the action under this transformation implies
\begin{equation}\label{Axc2}
\int \frac{\delta A_{xc}[n]}{\delta n({\bf r},t)} \vec \nabla_{{\bf r}} n({\bf r}) d{\bf r}  =0~,
\end{equation}
and an integration by parts leads to
\begin{equation}
\int n({\bf r}) \vec \nabla_{{\bf r}} \frac{\delta A_{xc}[n]}{\delta n({\bf r},t)} d{\bf r} =0~.
\end{equation}
This would be the zero-force theorem if we could identify $\frac{\delta A_{xc}[n]}{\delta n({\bf r},t)}$ with  $v_{xc}[n;{\bf r},t]$ which, of course, is incorrect.

Fortunately, the resolution of the puzzle is now at hand.  The point is that we are making {\it two} errors, which are luckily compensating each other,  leaving us with the correct result.  The first error is in Eq.~(\ref{Axc-invariance}):  it is not true that the xc action remains invariant under the transformation $n \to n'$.  The invariance of $A_{xc}$ was ``derived" in Ref.~(\cite{Vignale95}) by showing that the change of the internal action under this transformation depends only the density, not on the wave function, and therefore cancels out in the difference $A_0 - A_{0s}$.  However, we failed to include the boundary term $i \langle \psi_T[n]|\delta \psi_T[n]\rangle$, which {\it does} depend on the wave function and therefore does not cancel out, causing the xc action to vary, to first order in $n'-n$,  by 
\begin{equation}
\delta A_{xc} = i \langle \psi_T[n]|\delta \psi_T[n]\rangle -  i \langle \psi_{sT}[n]|\delta \psi_{sT}[n]\rangle~,
\end{equation}
where $|\delta \psi_T[n]\rangle \equiv |\psi_T[n']\rangle - |\psi_T[n]\rangle$ and $|\delta \psi_{sT}[n]\rangle \equiv |\psi_{sT}[n']\rangle - |\psi_{sT}[n]\rangle$.  
Therefore, Eq.~(\ref{Axc-invariance}) must be amended as follows:
\begin{eqnarray}\label{Axc2bis}
A_{xc}[n']=A_{xc}[n]+ i \langle \psi_T[n]|\delta \psi_T[n]\rangle -  i \langle \psi_{sT}[n]|\delta \psi_{sT},[n]\rangle~.
\end{eqnarray}
and Eq.~(\ref{Axc2}) is replaced by:
\begin{eqnarray}\label{Axc2bis}
\int \left\{\frac{\delta A_{xc}[n]}{\delta n({\bf r},t)}\right. - i \left.\left \langle \psi_T[n]\left\vert\frac{\delta \psi_T[n]}{\delta n({\bf r},t)}\right\rangle\right.+i \left \langle \psi_{sT}[n]\left\vert\frac{\delta \psi_{sT}[n]}{\delta n({\bf r},t)}\right\rangle\right.\right\} \vec \nabla_{{\bf r}} n({\bf r}) d{\bf r} =0~.
\end{eqnarray}
Integrating by parts, and using the correct formula for $v_{xc}$, Eq.~(\ref{mainresultxc}) we do indeed recover the zero force theorem $\int n({\bf r}) \vec \nabla_{{\bf r}} v_{xc}[n;{\bf r},t] d{\bf r} = 0$.

The lesson is quite general:  we are allowed to pretend that the $xc$ potential is the functional derivative of the action, provided we calculate that functional derivative incorrectly, i.e. ignoring the boundary contribution. This is exactly what we did (unwittingly)  in our earlier papers.

\section {Conclusion}
I believe that the foregoing analysis provides a straightforward and pedagogically transparent resolution of the causality paradox in TDDFT.  Compared to the resolutions proposed in Refs.~\cite{vanLeeuwen98,Mukamel05} the present approach is obviously much closer to the spirit of the original RG paper. Furthermore,  our approach allows us to understand why in many cases we can get correct results from an incorrect representation of the xc potential.  

In closing I wish to emphasize that what we have derived here is a {\it variational principle} for the time-dependent density.  A variational principle is not as strong as a minimum principle, yet it is strong enough to formulate a dynamical theory.  While the absolute numerical value of the RG action has no physical meaning  (because a multiplication of the wave function by an arbitrary phase factor changes its value by an arbitrary constant),  it must be borne in mind that the action determines the dynamics through its variations, and those variations are independent of the arbitrary additive constant (a similar situation occurs in classical mechanics, since the Lagrangian is defined up to an arbitrary total derivative with respect to time). 

%
 

\section{Acknowledgements}
I am very grateful to Ilya Tokatly for a critical reading of the manuscript and for suggesting the analysis of equal-time singularities in Appendix~\ref{ETS}, and to Carsten Ullrich for pressing a discussion of the adiabatic limit. This work has been supported by DOE under Grant No. DE-FG02-05ER46203.

\appendix \section{Equal time singularities in $\delta v[n;{\bf r},t]/\delta n({\bf r}',t')$}\label{ETS}
Following Tokatly~\cite{Tokatly05,Tokatly07} we write the exact local conservation laws for particle number and momentum:
\begin{equation}
\partial_t n+\nabla \cdot {\bf j}=0
\end{equation}
and
\begin{equation}
m\partial_t j_i +\partial_j\left(mnu_i u_j+P_{ij}\right)+n\partial_iv=0,
\end{equation}
where ${\bf j}$ is the current density, $j_i$ is its $i$-th cartesian component, $u_i=j_i/n$ is the velocity field, $\partial_j$ denotes the derivative with respect to $r_j$ (with implied summation over repeated indices), and finally  $P_{ij}$ is the stress tensor.   These equations are valid both for interacting and non-interacting systems and together define the time-dependent potential $v$ as a functional of the density, provided the velocity field and the stress tensor are regarded as functionals of the density.  Interaction effects enter implicitly  throughf the form of these functionals.

Taking the divergence of the second equation and making use of the first, we recast the system in the more explicit form 
\begin{equation}
\partial_i (n \partial_i v) =  m\partial_t^2 n +m\partial_i(u_i\partial_t n) -m\partial_i({\bf j} \cdot \nabla  u_i) -\partial_i\partial_jP_{ij}.
\end{equation}
This equation can be formally solved, yielding
\begin{equation}\label{vofn}
v =  \hat G \left[m\partial_t^2 n +m\partial_i(u_i\partial_t n) -m\partial_i({\bf j} \cdot \nabla  u_i) -\partial_i\partial_jP_{ij}\right],
\end{equation}
where $\hat G$ is the inverse of the operator $\partial_i n \partial_i$.

For the limited purpose of identifying the equal-time singularities in $\delta v/\delta n$ we can ignore any retardation in the functional dependence of  $P_{ij}$ and $u_i$ on the density.  Then the right hand side of Eq.~(\ref{vofn}) depends on the density and its first two derivatives $\partial_t n$ and $\partial_t^2 n$ at time $t$.  No higher derivatives are involved.
Then taking the functional derivative with respect to $n({\bf r}',t')$ we get a singularity proportional to $\ddot \delta(t-t')$ from  $\partial_t^2 n ({\bf r},t)$, a singularity proportional to $\dot \delta(t-t')$ from  $\partial_t n ({\bf r},t)$, and, of course, a singularity proportional to $\delta(t-t')$ from the terms that do not contain time-derivatives of the density. 

We can furthermore say that the coefficient of $\ddot \delta(t-t')$ is completely free of interaction effects, since the interactions enter only in the functional $P_{ij}$ and $u_i$.\footnote{It is worth noting that in a time-dependent {\it current} density functional theory, $u_i = j_i/n$ would not be a functional but a basic variable, so interaction effects would enter only through $P_{ij}$.}  And we observe that the $\dot \delta$ singularity vanishes in the linear response regime (small perturbations around the ground-state) because the current vanishes in the ground-state.  

Very little can be said in general about the explicit form of the $\delta$-function singularity.  In the linear response regime, the dependence of $P_{ij}$ on density has been extensively studied, but only in local or semi-local approximations.\cite{Tao07}  A fully nonlinear, but still local approximation to $P_{ij}$,  known as nonlinear  elastic local deformation approximation, has been formulated by Tokatly~\cite{Tokatly05} and studied by Ullrich and Tokatly~\cite{Ullrich06} in a model calculation.   An accessible review of this theory can be found in Chapter 8 of Ref.~\cite{TDFTBook}.

\section{Time-dependent position functional theory}\label{Model}
The evolution of electronic systems subjected to time-dependent potentials is in general too complicated to allow us to construct the functionals $v[n;{\bf r},t]$, $|\psi_T[n]\rangle$ etc.., even in the simplest non-trivial case of a two-electron system.  However, a simpler ``position-functional theory" can be easily formulated, which is conceptually equivalent to the full-fledged theory and allows us to demonstrate explicitly all the main points of the theory.

Our model is based on a two-particle system in one-dimension, with a time-dependent hamiltonian of the form
\begin{equation}\label{def-HF}
\hat H_F(t) = \frac{1}{2}\left[\hat p_1^2+\hat p_2^2+\hat x_1^2+\hat x_2^2\right]+\frac{k}{2}(\hat x_1-\hat x_2)^2 -F(t) \hat x_1,
\end{equation}
where $\hat x_1$, $\hat x_2$ are the position operators of the two particles $\hat p_1$ and $\hat p_2$ the canonical momentum operators, and $F(t)$ is a time-dependent force, which acts only on particle $1$.  The two particles are subjected to a parabolic potential well, and interact with each other with a harmonic force with ``elastic contant" $k>0$. The idea is that $\hat x_1$ plays the role of the density operator; its expectation value $x_1(t)$ is the time-dependent density, $-F(t)$ is the external potential.  As in TDDFT, one can show that the time-dependent position $x_1(t)$  and the initial state of the system at $t=0$ uniquely determine the force $f(t)$ that produces it; but in this case the functional $f[x_1;t]$ can be explicitly constructed.

For definiteness, we start from an initial state described by the wave function
\begin{equation}\label{def-psi0}
\psi_0(x_1,x_2)=Ce^{-X^2}e^{-\sqrt{1+2k}x^2/4}
\end{equation}
where $X\equiv(x_1+x_2)/2$, $x\equiv x_1-x_2$, and $C=\frac {(1+2k)^{1/8}}{\pi^{1/2}}$ is the normalization constant.  This is the ground-state of the hamiltonian for $F=0$.  The time evolution of this state under the full time-dependent hamiltonian is
\begin{eqnarray} \label{wavefunction}
\psi(x_1,x_2,t)&=&Ce^{i\phi(t)}e^{-[X-X_c(t)]^2}e^{2i [X-X_c(t)]{\dot X}_c(t)}\nonumber\\ &\times& e^{-\sqrt{1+2k}[x-x_c(t)]^2/4}e^{i[x-x_c(t)]{\dot x}_c(t)/2]}~,\nonumber\\
\end{eqnarray}
where $X_c(t)$ and $x_c(t)$ are the solutions of the classical equations of motion
\begin{eqnarray}
&&\ddot X_c(t)+X_c(t) = F(t)/2\nonumber\\
&&\ddot x_c(t)+(1+2k)x_c(t) = F(t)
\end{eqnarray}
with initial conditions $X_c(0)=\dot X_c(0) =0$ and  $x_c(0)=\dot x_c(0) =0$.  The phase factor $\phi(t)$ is the classical action (including the zero-point energy): 
\begin{equation}\label{def-phi}
\phi(t)=-\frac{1+\sqrt{1+2k}}{2} t+\int_0^t L(t')dt'~, 
\end{equation} 
where 
\begin{eqnarray}\label{Lagrangian}
L =\dot X_c^2-X_c^2+FX_c 
+\frac{\dot x_c^2}{4}- \left(\frac{1+2k}{4}\right)x_c^2+\frac{F}{2}x_c\nonumber\\
\end{eqnarray}
is the classical Lagrangian.

  The solution of the equations of motion is
\begin{eqnarray} \label{classicalsolutions}
X_c(t) &=&  \frac{1}{2}\int_0^t\sin(t-t')F(t')dt'\nonumber\\
x_c(t) &=&  \int_0^t\frac{\sin[\sqrt{1+2k}(t-t')]}{\sqrt{1+2k}} F(t')dt'~.
\end{eqnarray}

The solution of the quantum mechanical problem is obtained by substituting Eqs.~(\ref{classicalsolutions}) into Eqs.~(\ref{wavefunction}),~(\ref{def-phi}), and~(\ref{Lagrangian}).     It is immediately evident that $x_c(t)$ and $X_c(t)$ are the expectation values of the quantum mechanical center of mass operator $\hat X = (\hat x_1+\hat x_2)/2$ and relative position $\hat x = \hat x_1 - \hat x_2$ respectively.    The expectation values of $\hat x_1$ and $\hat x_2$ are given by
\begin{eqnarray}
\langle \psi(t)|\hat x_1|\psi(t)\rangle \equiv x_1(t) &=& X_c(t)+\frac{x_c(t)}{2}\nonumber\\
\langle \psi(t)|\hat x_2|\psi(t)\rangle \equiv x_2(t) &=& X_c(t)-\frac{x_c(t)}{2}~.
\end{eqnarray}

Our task is now to express the external force and the wave function as functionals of $x_1(t)$ -- the ``density" of our model.  To do this, we observe that $x_2(t)$ is related to $x_1(t)$ by the classical equation of motion
\begin{equation}
\ddot x_2(t)+x_2(t)=-k[x_2(t)-x_1(t)]
\end{equation}
with initial condition $x_2(0)=\dot x_2(0)=0$.  The solution of this equation, for given $x_1(t)$,  is
\begin{equation}\label{def-x2}
x_2(t) = k\int_0^t \frac{\sin[\sqrt{1+k} (t-t')]}{\sqrt{1+k}} x_1(t') dt'~.
\end{equation}
From this we can express  both $X_c(t)$ and $x_c(t)$ as functionals of $x_1(t)$, and hence the whole time-dependent wave function $\psi(x_1,x_2,t)$ as a functional of $x_1(t)$.  Furthermore, the force $f$, which produces the evolution $x_1(t)$ is given by $f(t)=\ddot x_1(t)+x_1(t)+k[x_1(t)-x_2(t)]$.  Upon substituting the functional dependence of $x_2(t)$ on $x_1(t')$ in the expression for $f(t)$ we obtain the force functional
\begin{eqnarray}\label{f-functional}
f[x_1;t]&=&\ddot x_1(t)+(1+k) x_1(t)\nonumber\\
&-& k ^2\int_0^t \frac{\sin[\sqrt{1+k} (t-t')]}{\sqrt{1+k}} x_1(t') dt'~.\nonumber\\
\end{eqnarray}
Observe how the force is uniquely and causally determined by $x_1(t)$.    Knowing the force we can construct the phase $\phi(t)$ (Eq.~(\ref{def-phi})) as a functional of $x_1$, by substituting $F=f[x_1]$ in the Lagrangian~(\ref{Lagrangian}). 
Finally, we construct the internal action functional
\begin{eqnarray}
A_0[x_1]&=&\int_0^T \langle \psi[x_1;t]|i\partial_t-\hat H_0|\psi[x_1,t]\rangle dt\nonumber\\
&=&-\int_0^T f[x_1;t']x_1(t') dt'~,
\end{eqnarray}
where $f[x_1,t]$ is given by Eq.~(\ref{f-functional}).  

We are now in a position to demonstrate explicitly the connection between the force and the functional derivative of the action.  Namely, we can prove that  
\begin{equation}\label{check1}
-f[x_1,t]=\frac{\delta A_0[x_1]}{\delta x_1(t)}- i \left\langle \psi_T[x_1] \left\vert \frac{\delta \psi_T[x_1]}{\delta x_1(t)}\right\rangle\right.~.
\end{equation}
where the state $|\psi_T[x_1]\rangle$  is described by the wave function~(\ref{wavefunction}), evaluated at time $T$ and expressed as a functional of $x_1[t]$ (the negative sign on the left hand side comes from the fact that  the force enters the hamiltonian $\hat H_F$ (Eq.~\ref{def-HF}) with a sign opposite to that of the potential in Eq.~(\ref{def-HV}).)  

The calculation is greatly simplified by the following two observations:  (i) All the terms that involve an expectation value of $\hat X-X_c$ or $\hat x - x_c$ are obviously zero, and (ii) because $x_c(t)$  and $X_c(t)$ are solutions of the classical equation of motion, the variation of the phase $\phi(T)$ comes only from the variation of the force (regarded as a functional of of $x_1$), and from the variation of $x_2(t)$ at the upper limit of integration.   In this way we easily arrive at
\begin{equation}
-i\left\langle \psi_T [x_1]\left\vert \frac{\delta \psi_T[x_1]}{\delta x_1(t)}\right\rangle\right.
=\int_0^T \frac{\delta f[x_1;t']}{\delta x_1(t)}x_1(t') dt' ~,
\end{equation}
from which Eq.~(\ref{check1}) follows at once.

Similarly, we can show that the ``Berry curvature"  $2\Im m  \left\langle \frac {\delta \psi_T[x_1]}{\delta x_1(t')} \left \vert  \frac {\delta \psi_T[x_1]}{\delta x_1(t)} \right\rangle \right.$ is given by
\begin{equation}
\frac{\delta x_2(T)}{\delta x_1(t')} \frac{\delta \dot x_2(T)}{\delta x_1(t)}-\frac{\delta x_2(T)}{\delta x_1(t)} \frac{\delta \dot x_2(T)}{\delta x_1(t')}, 
\end{equation}
so making use of Eq.~(\ref{def-x2}) we obtain
\begin{equation}
2\Im m  \left\langle \frac {\delta \psi_T[x_1]}{\delta x_1(t')} \left \vert  \frac {\delta \psi_T[x_1]}{\delta x_1(t)} \right\rangle \right.= k^2 \frac{\sin[\sqrt{1+k}(t-t')]} {\sqrt{1+k}}~,
\end{equation}
from which the arbitrary time $T$ has disappeared!  Armed with this result it is an easy matter to verify that
\begin{equation} \label{f-derivative}
-\frac{\delta f[x_1;t]}{\delta x_1(t')} = 2\Im m  \left\langle \frac {\delta \psi_T[x_1]}{\delta x_1(t')} \left \vert  \frac {\delta \psi_T[x_1]}{\delta x_1(t)} \right\rangle \right.~, 
\end{equation}
for $t>t'$, in agreement with Eq.~(\ref{result2}).  We can also verify the presence of the singularity at $t=t'$ discussed in section~\ref{section2} after Eq.~(\ref{result3}).  Indeed, the functional derivative of the first two terms in the expression of our force functional, Eq.~(\ref{f-functional}), gives
\begin{equation}\label{f-derivative-singular}
\left. \frac{\delta f[x_1;t]}{\delta x_1(t')}\right\vert_{sing} = \ddot \delta (t-t')+(1+k)\delta(t-t')~. 
\end{equation}
Notice that there is no term proportional to $\dot \delta$ in this simple model.

Finally, we observe that the analogue of the xc potential -- an xc force in this case --  is $F_{xc}[x_1;t] = f_s[x_1,t]-f[x_1,t]$, where the non-interacting force functional $f_s[x_1,t]$ is obtained from Eq.~(\ref{f-functional}) simply by putting $k=0$, so that
\begin{equation}\label{xcforce-s}
f_s[x_1,t] = \ddot x_1(t)+x_1(t)
\end{equation}
and 
\begin{eqnarray}\label{xcforce}
F_{xc}[x_1;t] &=& -kx_1(t)+k^2\int_0^t \frac{\sin [\sqrt{1+k}(t-t')]}{\sqrt{1+k}} x_1(t')\nonumber\\
&=&-k\{x_1(t)-x_2[x_1;t]\}.
\end{eqnarray}
Of course, $F_{xc}$  is nothing but the force exerted by the second particle on the first, expressed as a functional of the basic variable $x_1(t)$.    It is worth noting that the singular term $\ddot x_1(t)$ has cancelled out in $F_{xc}$. The singular part of the functional derivative of $F_{xc}$ is simply a $\delta$-function
\begin{equation}\label{fxc-derivative-singular}
\left. \frac{\delta F_{xc}[x_1;t]}{\delta x_1(t')}\right\vert_{sing} = -k\delta(t-t')~,
\end{equation}
in agreement with the discussion following Eq.~(\ref{result3-xc}).

Let us now demonstrate explicitly how, given the knowledge of the exact functional $F_{xc}[x_1;t]$ (Eq.~(\ref{xcforce})), one can calculate the evolution  of $x_1(t)$  within a ``Kohn-Sham  scheme". 
First of all, we introduce the ``Kohn-Sham hamiltonian"
\begin{equation}
\hat H_s(t) = \frac{1}{2}\left[\hat p_1^2+\hat p_2^2+\hat x_1^2+\hat x_2^2\right] -F(t) \hat x_1 -F_{xc}[x_1;t] \hat x_1(t)~,
\end{equation}
which describes two {non-interacting} particles, with an effective force $F+F_{xc}$ acting only on particle ``1". 
Then we solve the time-dependent Schr\"odinger equation $[i\partial_t - \hat H_s(t)] |\psi_s(t) \rangle =0$, starting with the non-interacting ground-state
\begin{equation}
\psi_{s0}(x_1,x_2) =  \frac{1}{\sqrt{\pi}} e^{-x_1^2/2}e^{-x_2^2/2}~,
\end{equation}
which clearly has the same expectation values of $\hat x_1$ and $\hat p_1$ as its interacting counterpart~(\ref{def-psi0}).  The solution is the time-dependent ``Kohn-Sham wave function"
\begin{eqnarray} \label{KSwavefunction}
\psi_s(x_1,x_2,t)&=&\frac{e^{i\phi_1(t)}}{\sqrt{\pi}} e^{-[x_1-x_{1c}(t)]^2/2}e^{i [x_1-x_{1c}(t)]{\dot x}_{1c}(t)}\nonumber\\ &\times& e^{-x_2^2/2}~,\nonumber\\
\end{eqnarray}
where $x_{1c}(t)$ is the solution of the equation of motion
\begin{equation}\label{eom-x1c}
\ddot x_{1c}(t)+x_1(t)= F(t)+F_{xc}[x_1;t] 
\end{equation}
with initial conditions $x_{1c}(0)=\dot x_{1c}(0)=0$,  and 
\begin{equation}
\phi_1(t)=-\frac{t}{2}+\int_0^t L_1(t') dt'~,
\end{equation}
with
\begin{equation}
L_1(t) = \frac{\dot x_{1c}^2}{2}-\frac{x_{1c}^2}{2} + (F(t)+F_{xc}[x_{1c};t])x_{1c}(t)~.
\end{equation}

Finally, we make use of the Kohn-Sham wave function to calculate the expectation value of $\hat x_1$:
\begin{equation}
\langle \psi_{s}(t)|\hat x_1|\psi_s(t)\rangle=x_{1c}(t)~.
\end{equation}

The equation of motion~(\ref{eom-x1c})  for $x_{1c}(t)$ can be rewritten explicitly as an integro-differential equation
\begin{eqnarray}
\ddot x_{1c}(t)+(1+k)x_{1c}(t) &&= F(t) \nonumber\\
&&+k^2\int_0^t \frac{\sin [\sqrt{1+k}(t-t')]}{\sqrt{1+k}} x_{1c}(t')~,\nonumber\\
\end{eqnarray}
with initial conditions $x_{1c}(0)=\dot x_{1c}(0)=0$.  This equation can be solved by Laplace transformation.  Denoting by $x_{1c}(s)$ the Laplace transform of $x_{1c}(t)$ we get
\begin{equation}
(s^2+1+k)x_{1c}(s) = F(s)+ \frac{k^2}{s^2+1+k}x_{1c}(s),
\end{equation}
and finally
\begin{eqnarray}
x_{1c}(s) &=&  \frac{s^2+1+k}{(s^2+1)(s^2+1+2k)}F(s)\nonumber\\
&=& \left[\frac{1}{s^2+1}+\frac{1}{s^2+1+2k}\right]\frac{F(s)}{2}~.
\end{eqnarray}
Going back to the time domain we finally obtain
\begin{eqnarray}
x_{1c}(t)&=& \frac{1}{2}\int_0^t \sin(t-t')F(t')\nonumber\\
&+&\frac{1}{2}\int_0^t \frac{\sin [\sqrt{1+2k}(t-t')]}{\sqrt{1+2k}}F(t')~,
\end{eqnarray}
which of course agrees with the exact solution $x_1(t)=X_c(t)+x_c(t)/2$ obtained from Eqs.~(\ref{classicalsolutions}).

In TDDFT  we do not have the luxury of knowing the exact xc force functional. But this example shows that,  if we knew it, we could use it to predict the exact evolution of the density.  So there is every reason to believe that a good approximation to the exact xc potential, obtained by whatever means, would enable us to make good predictions for the time evolution of the density.

\end{document}